\documentclass{elsart}

\usepackage[square,comma]{natbib}
\usepackage{graphicx}
\usepackage{pxfonts}
\usepackage{lineno}

\usepackage{amssymb}

\journal{}

\begin{document}

\thispagestyle{empty}
\begin{Large}
\textbf{DEUTSCHES ELEKTRONEN-SYNCHROTRON}

\textbf{\large{Ein Forschungszentrum der
Helmholtz-Gemeinschaft}\\}
\end{Large}

DESY 10-004

January 2010

\begin{eqnarray}
\nonumber &&\cr \nonumber && \cr \nonumber &&\cr
\end{eqnarray}
\begin{eqnarray}
\nonumber
\end{eqnarray}
\begin{center}
\begin{Large}
\textbf{Scheme for femtosecond-resolution pump-probe experiments
at XFELs with two-color ten GW-level X-ray pulses}
\end{Large}
\begin{eqnarray}
\nonumber &&\cr \nonumber && \cr
\end{eqnarray}

\begin{large}
Gianluca Geloni,
\end{large}
\textsl{\\European XFEL GmbH, Hamburg}
\begin{large}

Vitali Kocharyan and Evgeni Saldin
\end{large}
\textsl{\\Deutsches Elektronen-Synchrotron DESY, Hamburg}
\begin{eqnarray}
\nonumber
\end{eqnarray}
\begin{eqnarray}
\nonumber
\end{eqnarray}
ISSN 0418-9833
\begin{eqnarray}
\nonumber
\end{eqnarray}
\begin{large}
\textbf{NOTKESTRASSE 85 - 22607 HAMBURG}
\end{large}
\end{center}
\clearpage
\newpage

\begin{frontmatter}



\title{Scheme for femtosecond-resolution pump-probe experiments at XFELs with two-color ten GW-level X-ray pulses}


\author[XFEL]{Gianluca Geloni\thanksref{corr},}
\thanks[corr]{Corresponding Author. E-mail address: gianluca.geloni@xfel.eu}
\author[DESY]{Vitali Kocharyan}
\author[DESY]{and Evgeni Saldin}

\address[XFEL]{European XFEL GmbH, Hamburg, Germany}
\address[DESY]{Deutsches Elektronen-Synchrotron (DESY), Hamburg,
Germany}

\begin{abstract}
This paper describes a scheme for pump-probe experiments that can
be performed at LCLS and at the European XFEL and determines what
additional hardware development will be required to bring these
experiments to fruition.  It is proposed to derive both pump and
probe pulses from the same electron bunch, but from different
parts of the tunable-gap baseline undulator. This eliminates the
need for synchronization and cancels jitter problems. The method
has the further advantage to make a wide frequency range
accessible at high peak-power and high repetition-rate.  An
important feature of the proposed scheme is that the hardware
requirement is minimal. Our technique is based in essence on the
"fresh" bunch technique. For its implementation it is sufficient
to substitute a single undulator module with short magnetic delay
line, i.e. a weak magnetic chicane, which delays the electron
bunch with respect to the SASE pulse of half of the bunch length
in the linear stage of amplification. This installation does not
perturb the baseline mode of operation. We present a feasibility
study and we make exemplifications with the parameters of the
SASE2 line of the European XFEL.
\end{abstract}

%
%

\end{frontmatter}



\section{\label{sec:intro} Introduction}

With the spectacular result obtained at LCLS \cite{LCLS1,LCLS2}
x-ray free-electron lasers have become a reality. This achievement
relies on a high-performance beam formation system, which works as
in the ideal operation scenario described in the conceptual design
report \cite{LCLS1}. In particular, the small electron-beam
emittance achieved ($0.4 \mu$m with $0.25$ nC charge) allows
saturation within $20$ undulator cells, out of the $33$ available.

\begin{table}
\caption{Parameters for the short pulse mode used in this paper.
The undulator parameters are the same of those for the European
XFEL, SASE2, at 17.5 GeV electron energy.}

\begin{small}\begin{tabular}{ l c c}
\hline
& ~ Units &  Short pulse mode \\
\hline
Undulator period      & mm                  & 47.9   \\
Undulator length      & m                   & 256.2  \\
Length of undulator segment        & m                   & 5.0    \\
Length of intersection             & m                   & 1.1    \\
Total number of undulator cells    & -                   & 42     \\
K parameter (rms)     & -                   & 2.513  \\
$\beta$               & m                   & 17     \\
Wavelength            & nm                  & 0.15   \\
Energy                & GeV                 & 17.5   \\
Charge                & nC                  & 0.025  \\
Bunch length (rms)    & $\mu$m              & 1.0    \\
Normalized emittance  & mm~mrad             & 0.4    \\
Energy spread         & MeV                 & 1.5    \\
\hline
\end{tabular}\end{small}
\label{tab:fel-par}
\end{table}
\begin{figure}[tb]
\includegraphics[width=0.5\textwidth]{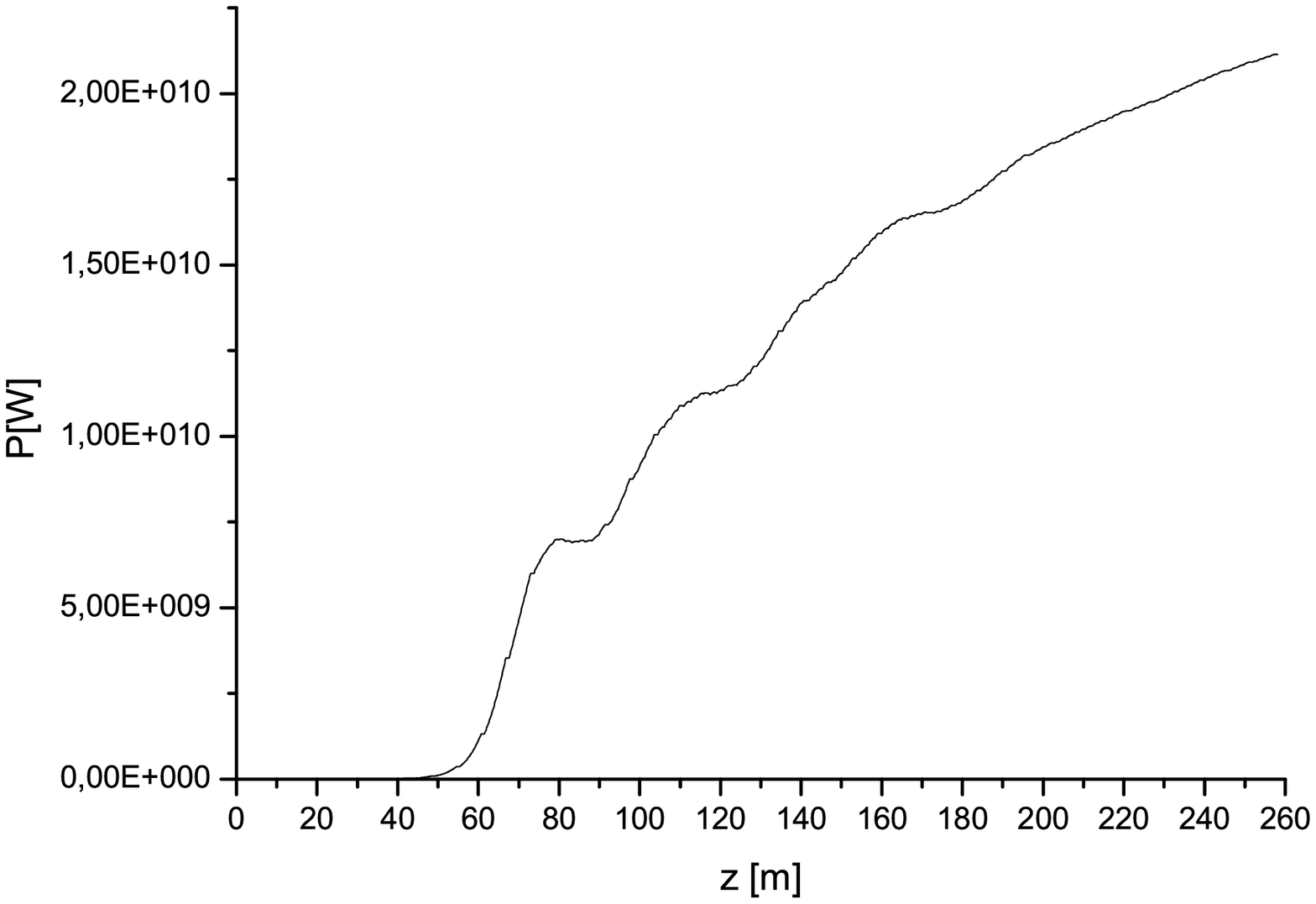}
\includegraphics[width=0.5\textwidth]{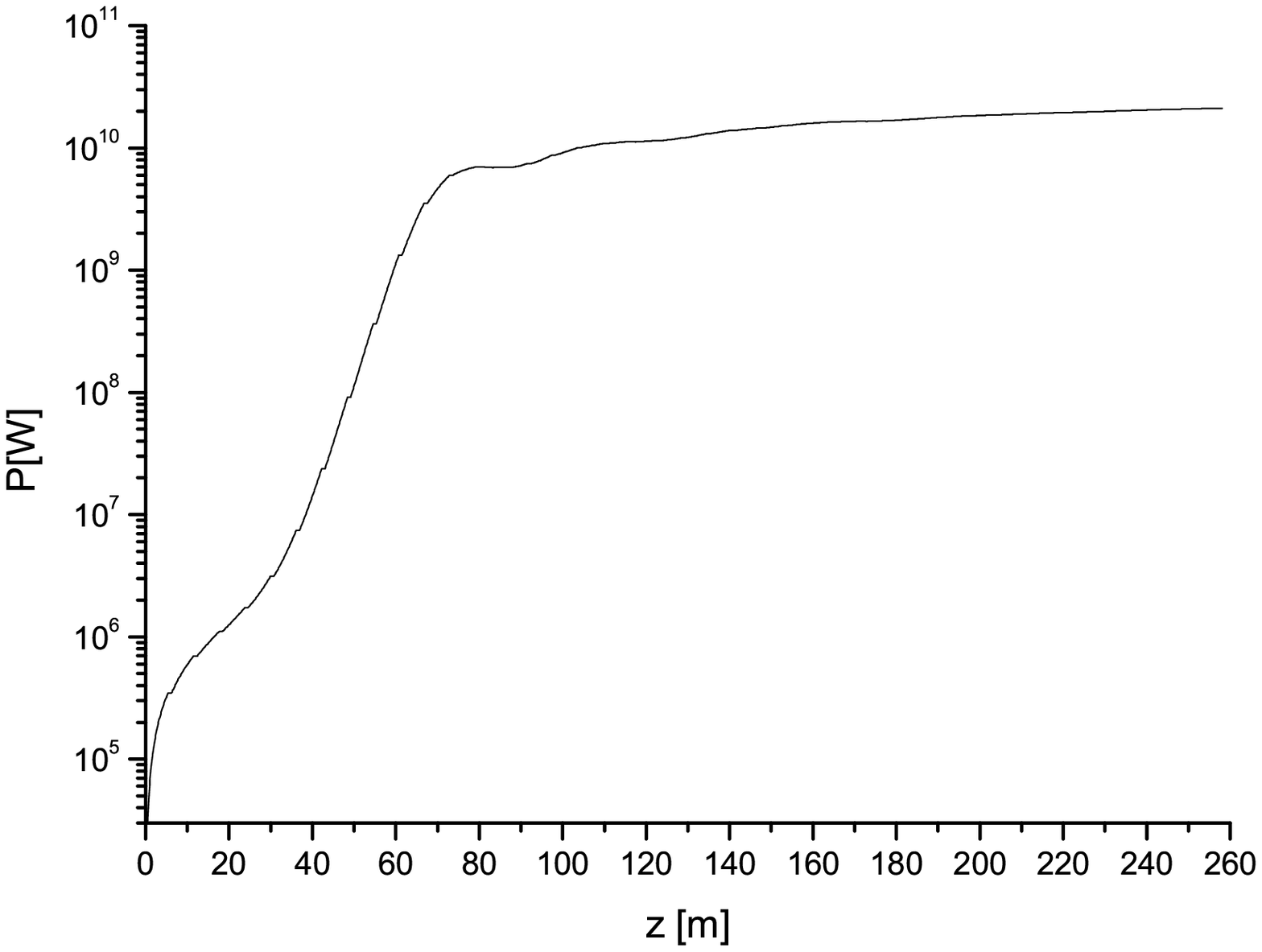}
\caption{Averaged FEL output power from the baseline setup. Left,
in linear scale. Right, in logarithmic scale.} \label{sase2base}
\end{figure}
\begin{figure}[tb]
\includegraphics[width=0.5\textwidth]{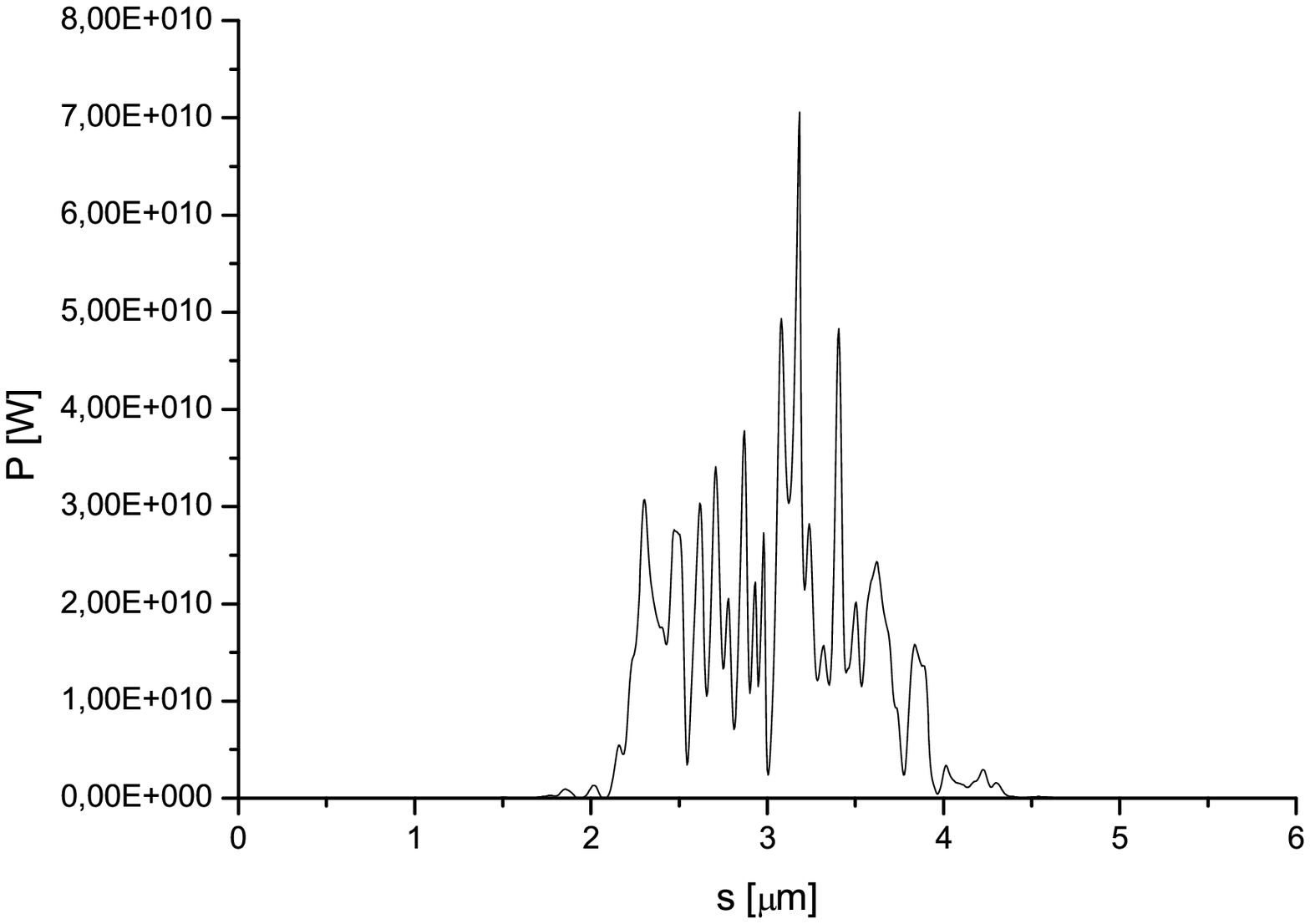}
\includegraphics[width=0.5\textwidth]{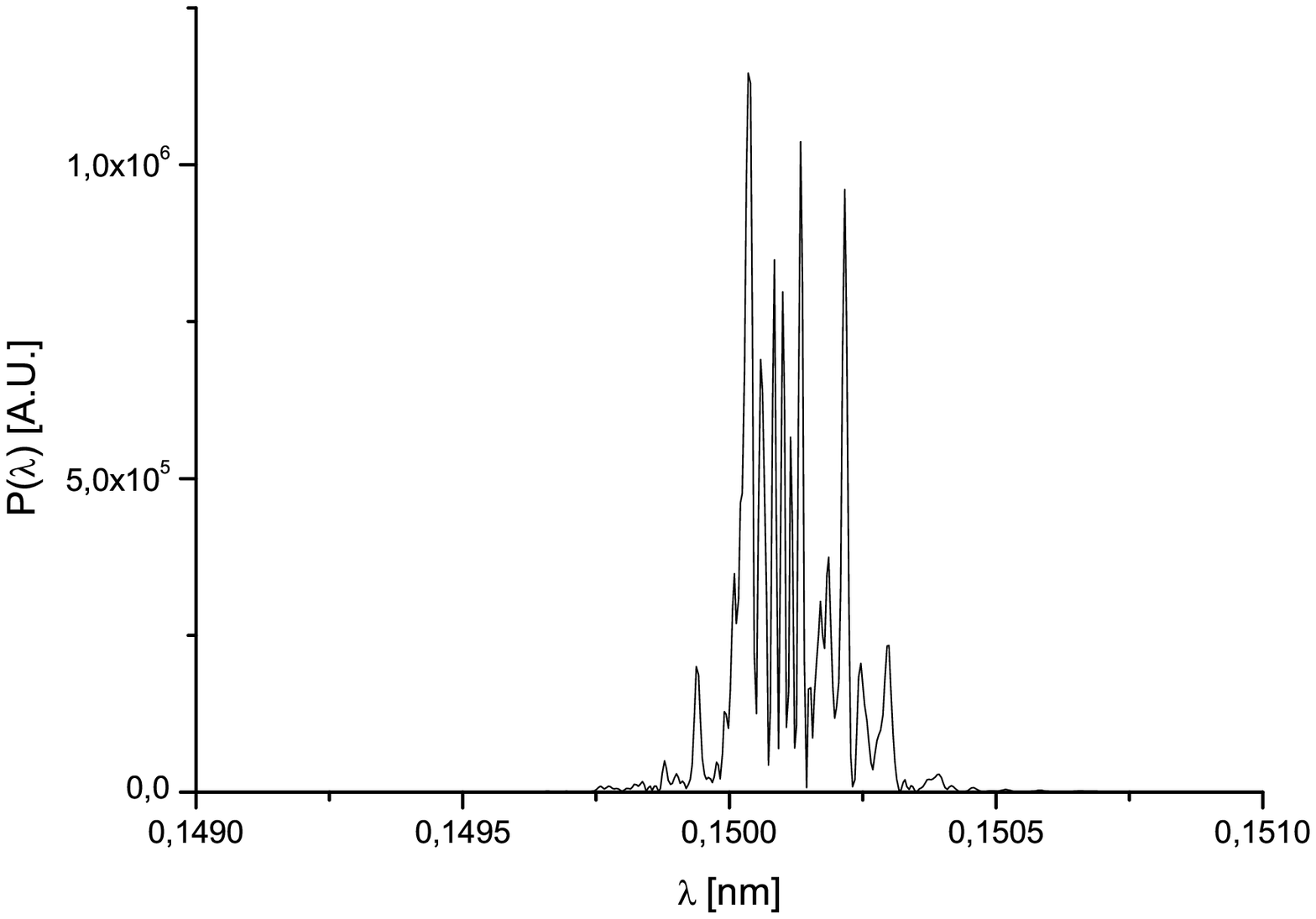}
\caption{Beam power distribution (left) and spectrum (right) after
12 undulator segments ($73.2$ m).} \label{sase2psat}
\end{figure}
\begin{figure}[tb]
\includegraphics[width=0.5\textwidth]{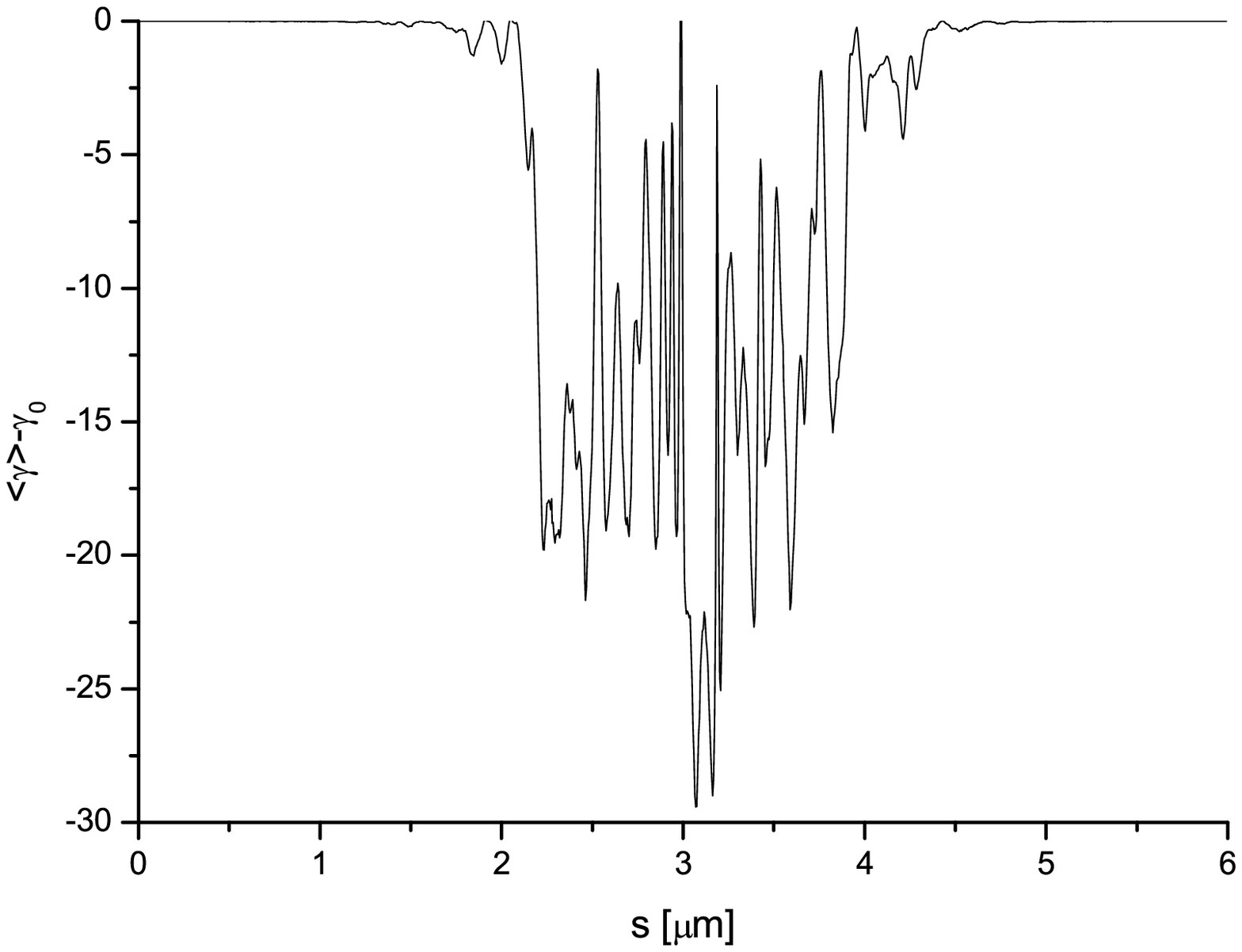}
\includegraphics[width=0.5\textwidth]{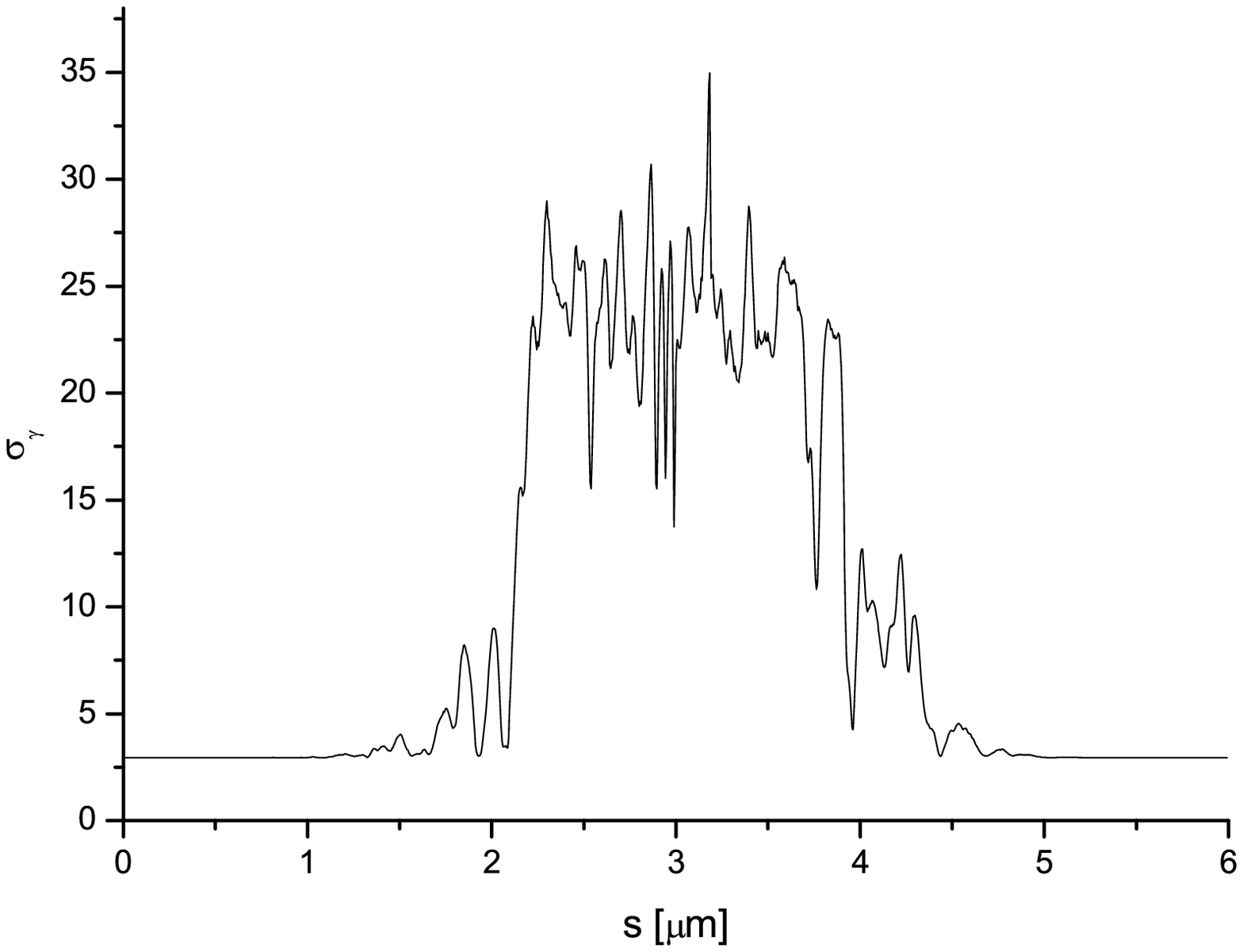}
\caption{Electron beam energy loss (left) and induced energy
spread (right) after 12 undulator segments ($73.2$ m).}
\label{sase2en}
\end{figure}
Besides the baseline mode, LCLS can also work in a low charge mode
\cite{DING} ($Q=30$ pC), enabling even smaller emittance (as
$\epsilon \sim \sqrt{Q}$), with the same peak current $I_{peak}$.
Since space-charge effects are reduced, the low charge allows for
a more efficient compression i.e. for shorter electron bunches
and, therefore, for shorter radiation pulses. The method described
in \cite{DING} allows for the production of ten GW- level, sub 10
fs, coherent X-ray pulses.

The production of short, pulses of high-power, coherent x-ray
radiation are of great importance when it comes to time-resolved
experiments, which are used to monitor time-dependent phenomena.
In a typical pump-probe experiment, a short probe pulse follows a
short pump pulse at some specified delay. Femtosecond capabilities
have been available for some years at visible wavelengths.
However, there is a strong interest in extending pump-probe
techniques to X-ray wavelengths because they allow to probe
directly structural changes with atomic resolution. One of the
main technical problems for building pump-probe capabilities is
the synchronization between pump and probe pulses, which should
have different colors.

Here we propose a method to get around this obstacle by using a
two step FEL process, in which two different frequencies (colors)
are generated by the same femtosecond electron bunch
\cite{FELD,GENS}. It has the further advantage to make a wide
frequency range accessible at high peak power and high repetition
rate. Our technique is based in essence on the fresh bunch
technique \cite{HUAYU,SAL1,SAL2}, and exploits the short
saturation length experimentally demonstrated at LCLS. We consider
in particular the case for the low charge, short-pulse mode of
operation. Such mode of operation can be successfully taken
advantage of at the European XFEL as well. In our analysis we
consider baseline parameters similar to those of the SASE2 line at
the European XFEL, so that our scheme can be implemented at the
very first stage of operation of this facility. Table
\ref{tab:fel-par} reports these parameters. Fig. \ref{sase2base},
Fig. \ref{sase2psat} and Fig. \ref{sase2en} further exemplify the
short-pulse mode of operation showing respectively the average
output power, the power and spectrum at saturation, and the energy
spread and energy loss of the electron bunch, also at saturation.
It should be remarked that the applicability of our method is
obviously not restricted to the European XFEL setup. Other
facilities e.g. LCLS may benefit from this work as well.
\begin{figure}
\begin{center}
\includegraphics*[width=100mm]{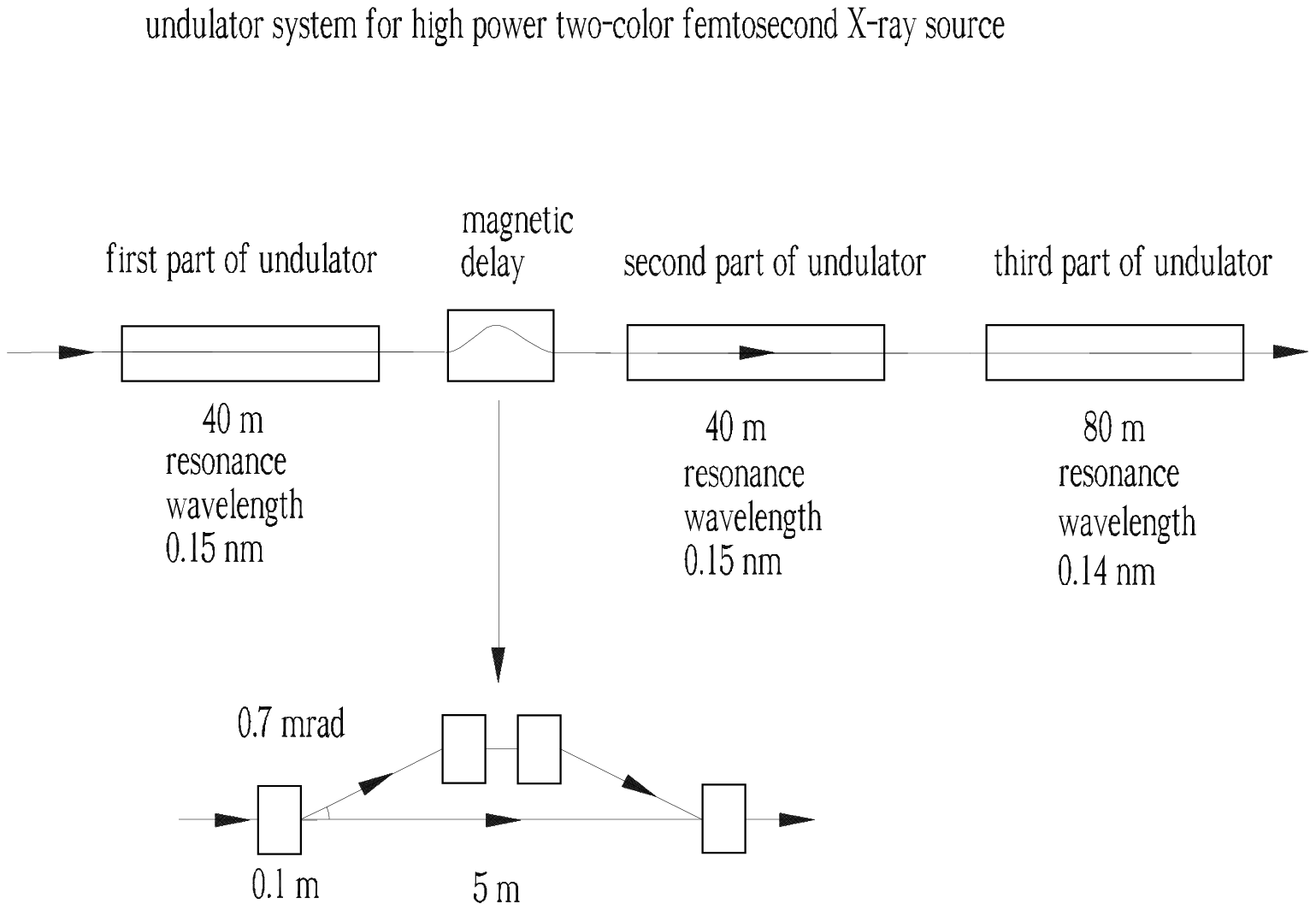}
\caption{\label{undusys} Design of the undulator system for
two-color femtosecond X-ray source.  }
\end{center}
\end{figure}
As we will discuss in Section \ref{long}, our method can also be
used in the case of long-pulse mode of operation. In fact, there
is no principle difference between long-pulse and short-pulse
modes of operation concerning our pump-probe technique.  However,
in the case of short-pulse mode of operation, described in the
next Section \ref{scheme}, the hardware requirement is minimal,
and through this paper we will mainly focus on this case. For its
implementation it is sufficient to substitute a single undulator
segment with a short magnetic chicane, Fig. \ref{undusys}, whose
function is both, to wash out the electron bunch microbunching,
and to delay the electron bunch with respect to the x-ray pulse
produced, in the linear regime, in the first part of the
undulator. In this way, half of the electron bunch is seeded, and
saturates in the second part of the undulator. Finally, the second
half of the electron bunch, which remains unspoiled, lases in the
third part of the undulator at a different wavelength.

In the following we present a feasibility study of the method
(section \ref{scheme}), and a few further developments (section
\ref{reldelay} and \ref{long}), before coming to conclusions.

\section{Feasibility study\label{scheme}}

\begin{figure}[tb]
\begin{center}
\includegraphics[width=100mm]{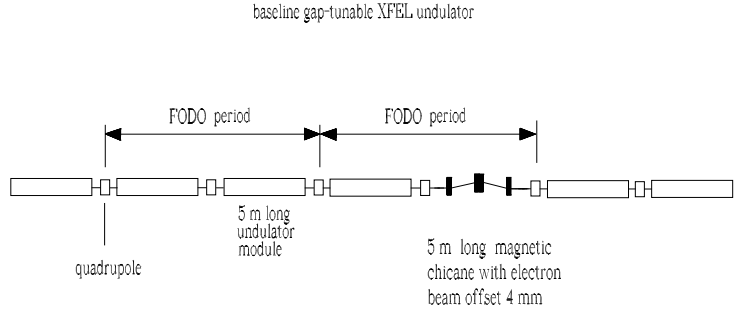}
\caption{\label{delaymag} Installation of a magnetic delay in the
baseline XFEL undulator.}
\end{center}
\end{figure}
With a baseline gap-tunable undulator design (like in the SASE2
case) the pump-probe option in the short-pulse mode of operation
only requires the installation of a magnetic delay. Thus, the
hardware required for the implementation of our scheme is minimal,
and consists in the substitution of one of the undulator segments
with a magnetic chicane, as shown in Fig. \ref{undusys}. The
quadrupole separation of the undulator FODO lattice ($6.1$ m) is
large enough so that a relatively short ($5$ m) magnetic chicane
can be installed (see Fig. \ref{delaymag}).
\begin{figure}
\begin{center}
\includegraphics*[width=100mm]{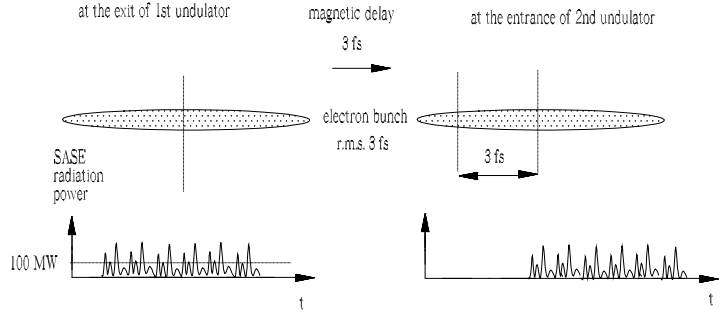}
\caption{\label{freshort} Sketch of principle of "fresh bunch"
technique for short (6 fs) pulse mode operation.  }
\end{center}
\end{figure}
The electron beam first goes through the first undulator as in the
baseline design, producing SASE radiation in the linear regime.
After the first undulator the electron beam is guided through the
magnetic chicane. The trajectory of the electron beam in the
chicane has the shape of an isosceles triangle with the base equal
to the undulator segment length, $L_w$. The angle adjacent to the
base, $\theta$, is considered to be small. The magnetic delay
needed for the generation of two color x-ray pulses should satisfy
three requirements.

First, the radiation pulse must overlap only half of the electron
bunch at the chicane exit, i.e. the electron beam extra path
length must be of the order of the rms of electron bunch length,
as shown in Fig. \ref{freshort}. Second, collective effects,
namely coherent synchrotron radiation (CSR) should be avoided in
order to preserve transverse emittance. In the present case,
simple estimations show that CSR  should not be a serious
limitation in our case. Third, the electron beam modulation
introduced in the first undulator due to FEL interaction must be
washed out. The presence of a local energy spread in the electron
beam naturally solves the problem. In fact, for Gaussian local
energy spread, and neglecting collective effects, the amplitude of
the density modulation $a$ at the chicane exit is given by


\begin{eqnarray}
a = a_0 \exp\left[-\frac{1}{2}\frac{\left\langle(\delta
\gamma)^2\right\rangle}{\gamma^2}\frac{R_{56}^2
}{\lambdabar^2}\right] \label{uno}
\end{eqnarray}
where $a_0$ is the amplitude of the density modulation at the
entrance of the chicane, $R_{56}$ is the momentum compaction
factor and $\lambdabar$ is the reduced wavelength. In our case,
parameters of interest are the undulator segment length $L_w = 5$
m, the deflection angle $\theta = 0.7$ mrad, the dispersion
$R_{56} = L_w \theta^2 \sim 2.5 \mu$m, and the relative energy
spread $\Delta \gamma/\gamma \sim 0.01 \%$, corresponding to an
energy spread of about $1.5$ MeV. This leads to the suppression of
the beam modulation by a factor of about $\exp(-20)$. Here we
assume that an uncorrelated relative energy spread of $1.5$ MeV is
already present at the
entrance of the SASE undulator. 

One may also account for the linear energy chirp in the electron
beam by introducing a resonance frequency-shift in the second
undulator. Thus, we can neglect the linear energy chirp and
account for the non-linear energy chirp only. We further require
that the non-linear energy chirp be sufficiently small, and that
the energy deviation across the lasing part of the bunch be
smaller than the FEL $\rho$ parameter. In our case study $\rho
\sim 0.1 \%$, and simple estimations show that the energy chirp
should not be a serious limitation.

After the chicane, the relative positions of radiation and
electrons at the entrance of the second undulator part looks like
on the right of Fig. \ref{freshort}. The left half of the bunch
will start the usual SASE process form shot noise. The right half,
instead, is seeded by the radiation produced in the first part of
the undulator, and will reach saturation much sooner. Finally, a
third undulator part, tuned at a slightly different frequency,
will allow two-color operation: the right half of the bunch is
spoiled by the SASE process and cannot radiate anymore, while the
left half emits usual SASE radiation up to saturation.  Low-charge
bunch simulations are not time expensive, and we it find
convenient to use the Genesis code \cite{GENE}, i.e. the same code
used for simulations at LCLS. In the following we describe the
outcomes our of these computer simulations.

\subsection{First stage \label{First}}

\begin{figure}[tb]
\includegraphics[width=1.0\textwidth]{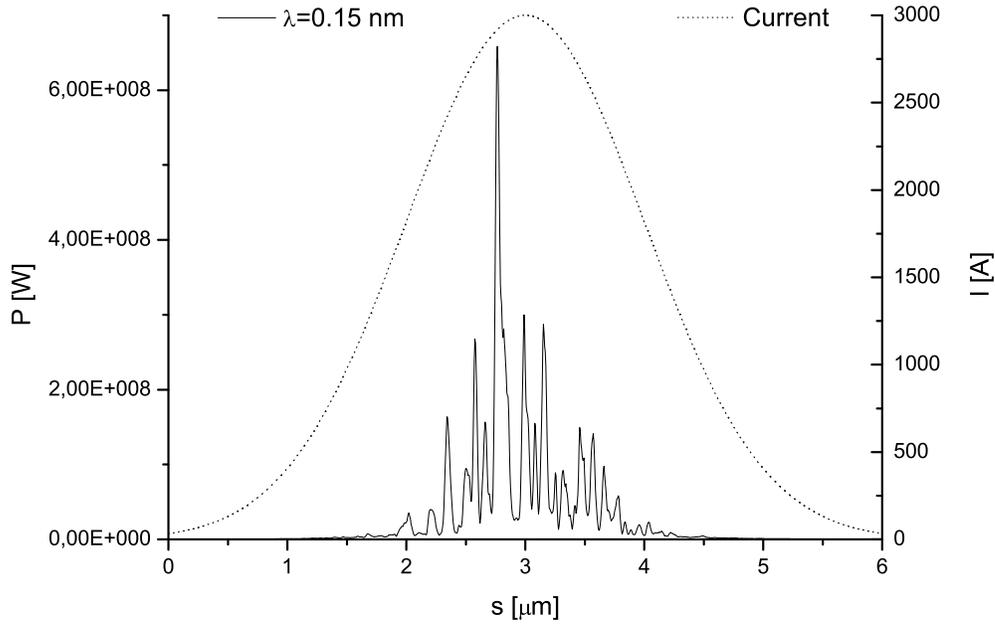}
\caption{Beam power distribution at the end of the first stage
after 7 cells ($42.7$ m). The bunch current is also shown.}
\label{IstageP}
\end{figure}
\begin{figure}[tb]
\includegraphics[width=0.5\textwidth]{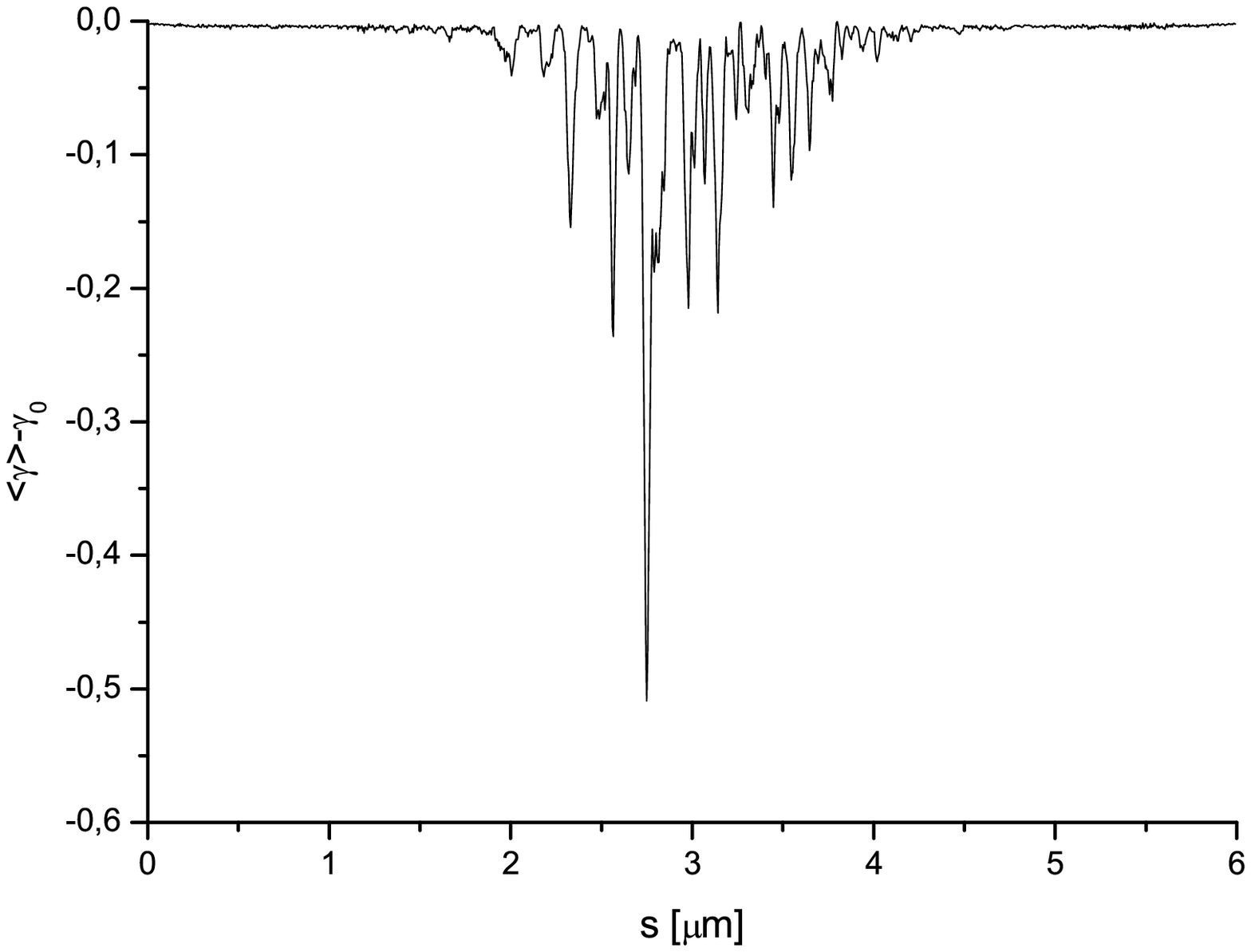}
\includegraphics[width=0.5\textwidth]{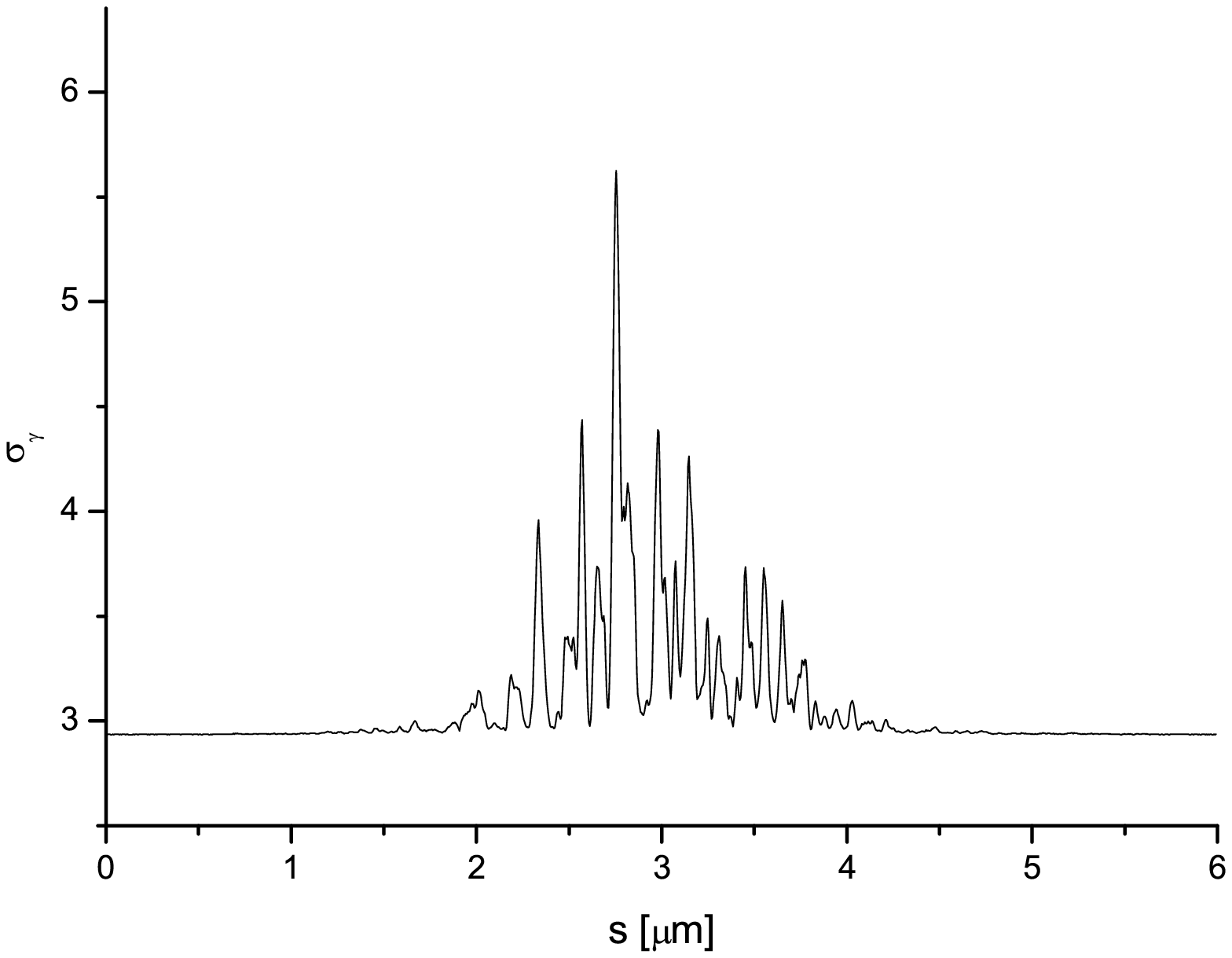}
\caption{Electron beam energy loss (left) and induced energy
spread (right) at the end of the first stage after 7 cells ($42.7$
m).} \label{Istageen}
\end{figure}
First we performed a simulation in the SASE mode for the first $7$
cells (each consisting of a $5$m-long undulator segment and a
$1.1$m-long intersection), for a total length of $42.7$ m. The
beam power distribution is shown in Fig. \ref{IstageP}. The
electron beam energy loss and the induced energy spread are shown
in Fig. \ref{Istageen}.

The FEL process is still in the linear regime, and the induced
energy spread and energy loss are small (see Fig. \ref{Istageen}),
compared to the values at saturation. Nevertheless, $7$ cells are
sufficient to yield about $100$ MW output (see Fig.
\ref{IstageP}), to be used as a seed in the second part of the
undulator.

\subsection{Magnetic delay \label{delay}}

\begin{figure}[tb]
\includegraphics[width=1.0\textwidth]{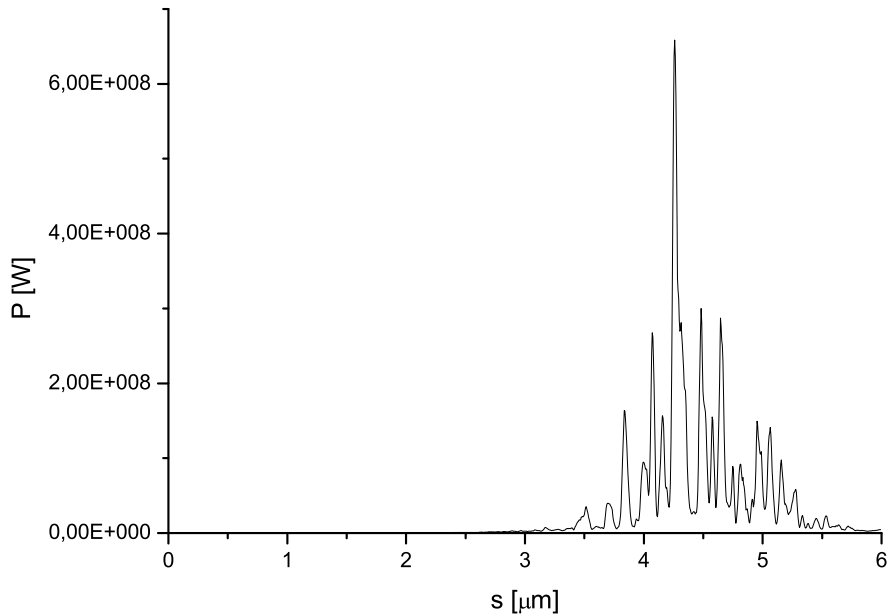}
\caption{Beam power distribution after the magnetic delay. }
\label{IstageS}
\end{figure}

Following the first undulator part, one has to model the magnetic
delay. This is accounted for by shifting on the right the output
field from the first part of the undulator\footnote{For this
particular simulation example, radiation was shifted right of
$1.5~\mu$m.}, thus simulating a relative delay of the electron
bunch. The field distribution obtained in this way, which is fed
into the simulations for the second undulator part, is shown in
Fig. \ref{IstageS}.

It should be noted that the energy modulation of the electron
bunch is suppressed in the same degree as the density modulation
(aside for a different pre-exponential factor).  Therefore, we can
conclude that the problem of suppressing the beam modulation
induced in the first undulator is solved quite naturally due to
the presence of energy spread in the electron beam\footnote{Here
we assume a Gaussian uncorrelated energy spread at the entrance of
the first undulator of about  $0.01\%$. This spread can be
introduced simply by tuning the gap of the first $3$ modules of
the first undulator  to $0.4 nm$. In this way, the magnetic field
is large enough to guarantee generation of the energy-spread level
needed due to quantum diffusion.}. As a result, the initial
modulation of the electron beam at the entrance of the second
undulator is given by shot noise only. It should also be remarked
here that we use a relatively weak chicane for which the spread of
the phase shift between electrons due to energy spread (magnetic
delay is non isochronous) is a few wavelengths only. This is more
than sufficient for suppressing (smearing) the short
radiation-wavelength scale modulation into electron beam, but does
not change the large FEL coherence-length scale modulation
presented in Fig. \ref{Istageen}.

\subsection{Second stage \label{second}}

\begin{figure}[tb]
\includegraphics[width=1.0\textwidth]{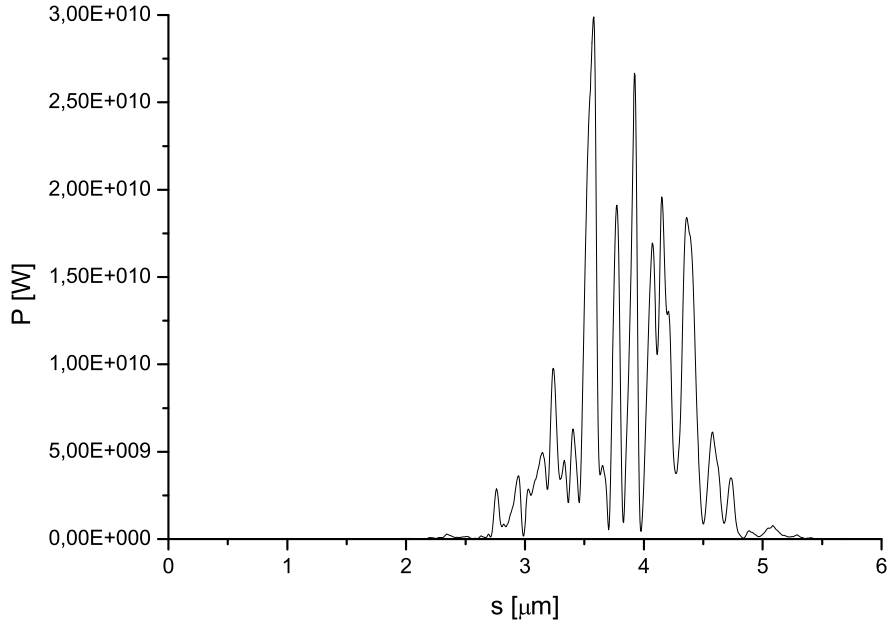}
\caption{Beam power distribution at the end of the second stage,
$7$ cells-long ($42.7$ m).} \label{IIstageP}
\end{figure}

\begin{figure}[tb]
\includegraphics[width=0.5\textwidth]{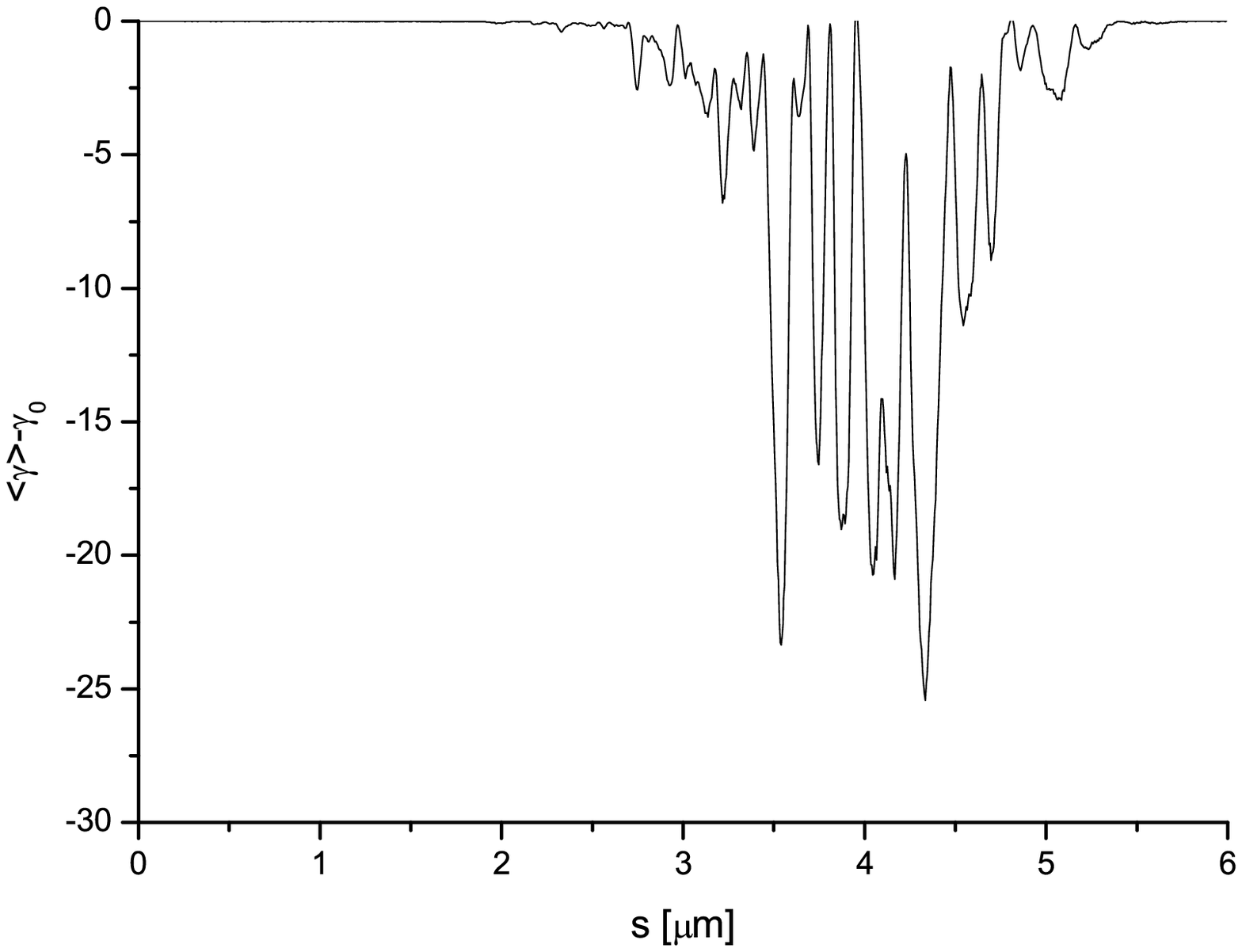}
\includegraphics[width=0.5\textwidth]{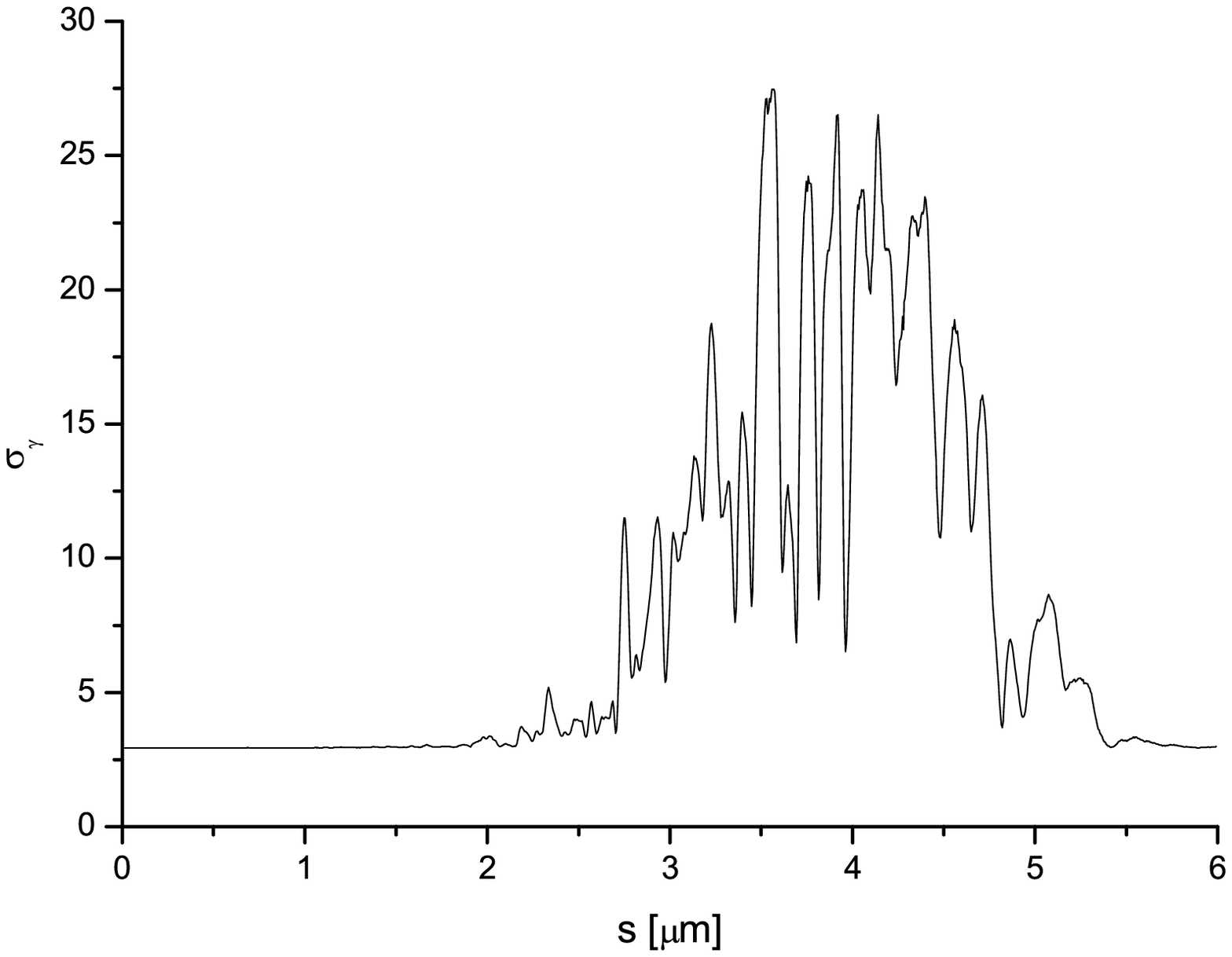}
\caption{Electron beam energy loss (left) and induced energy
spread (right) at the end of the second stage, $7$ cells-long
($42.7$ m).} \label{IIstageen}
\end{figure}

Together with the shifted radiation pulse, also an electron beam
generated using the values of energy loss and energy spread at the
exit of the first stage is fed in the simulation of the second
undulator part. The second undulator part is taken to be another
$7$ cells long. The output power distribution is shown in Fig.
\ref{IIstageP}, while energy loss and energy spread are plotted in
Fig. \ref{IIstageen}.

As one can see, the seeded part of the electron bunch reaches
saturation with ten GW power level, of the order of the baseline
operation mode (see Fig. \ref{sase2psat}). The left part of the
electron bunch produces SASE radiation in the linear regime only,
which is negligible. The correctness of these outcomes is also
supported by inspection of Fig. \ref{IIstageen} and comparison
with Fig. \ref{sase2en}. The fact that the right part of the
electron bunch is spoiled is evident from such comparison. The
left part of the electron bunch has effectively not lased yet, and
can be further used in the third undulator part.

\subsection{Third stage \label{third}}

\begin{figure}[tb]
\includegraphics[width=1.0\textwidth]{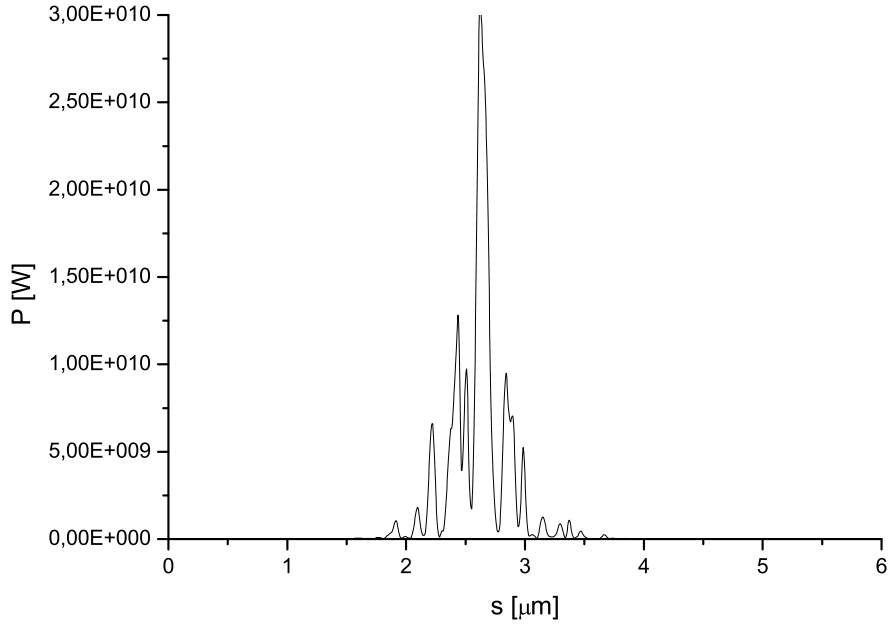}
\caption{Beam power distribution at the end of the third stage,
$11$ cells-long ($67.1$ m).} \label{IIIstageP}
\end{figure}

\begin{figure}[tb]
\includegraphics[width=0.5\textwidth]{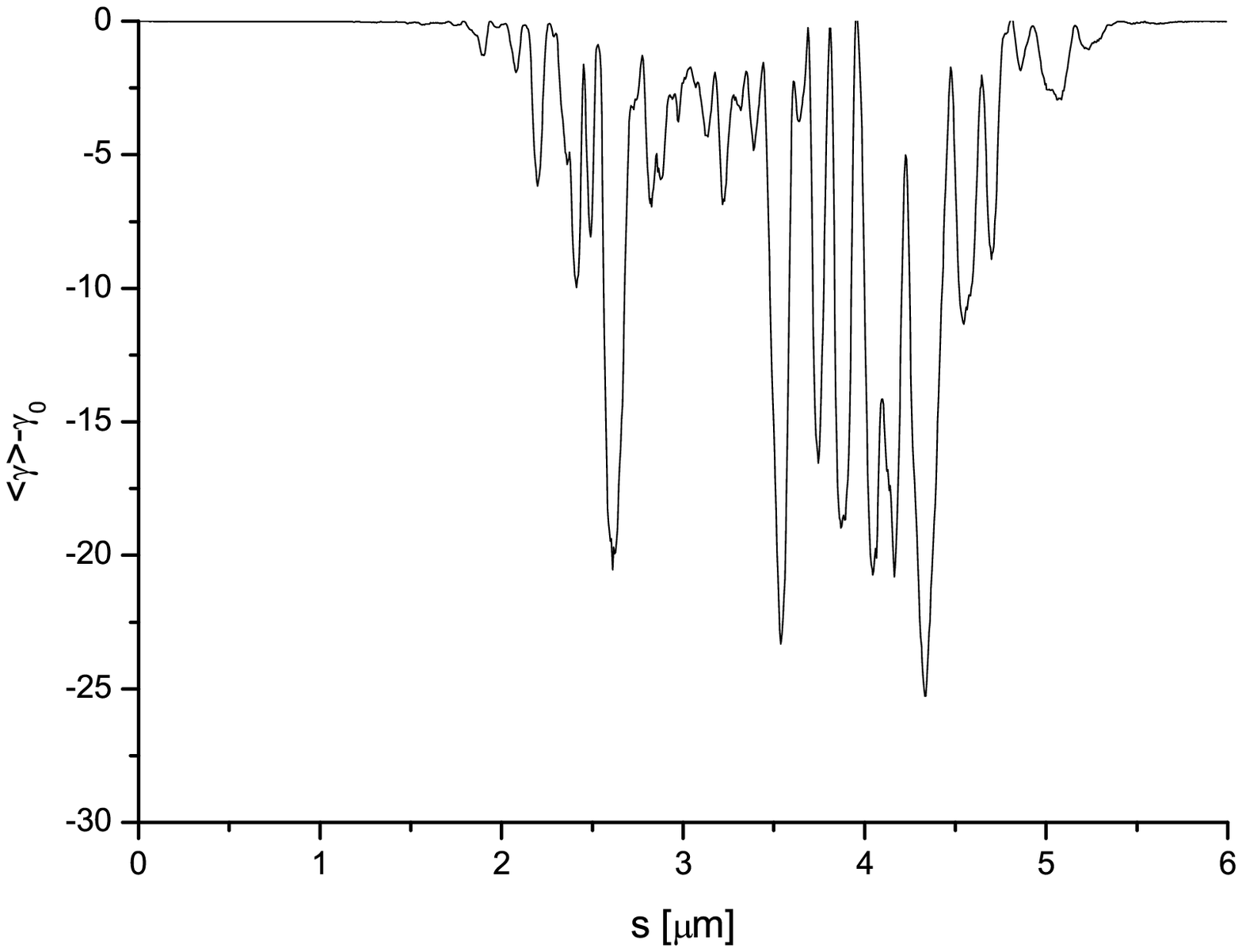}
\includegraphics[width=0.5\textwidth]{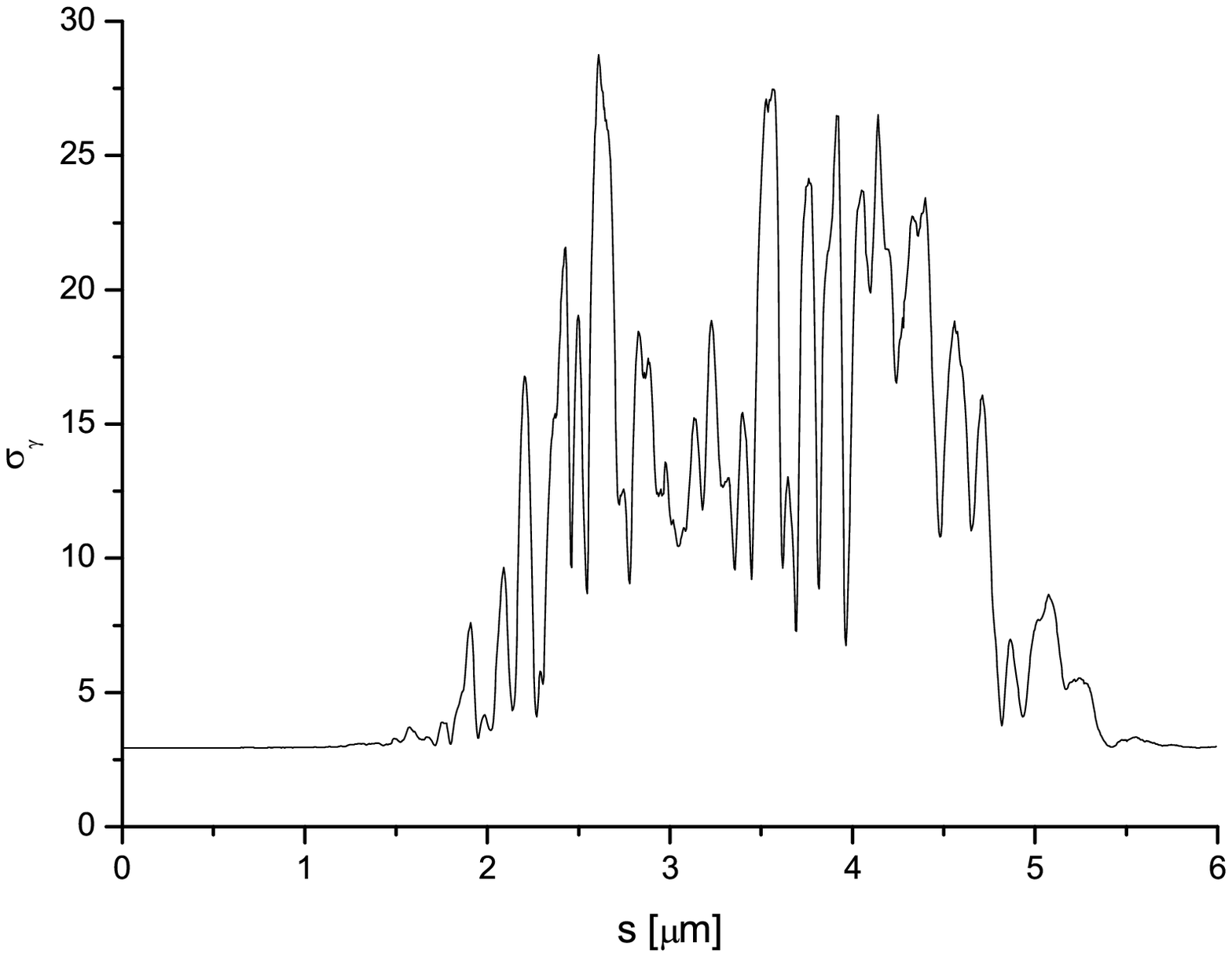}
\caption{Electron beam energy loss (left) and induced energy
spread (right) at the end of the third stage $11$ cells-long
($67.1$ m).} \label{IIIstageen}
\end{figure}

The simulations for the third undulator part consist once more of
SASE simulations, where the electron beam used is generated, as
described in section \ref{second}, using the values of energy loss
and energy spread at the end of the second undulator part. The
resonant wavelength is now set, for exemplification purposes, to
$0.14$ nm. Obviously, such choice is arbitrary\footnote{Within the
tunability range of SASE2, and within the available length for the
second stage ($160$ m) it is possible to reach saturation at least
within full range 0.1-0.4 nm.}. The output power distribution
after further $11$ undulator segments is shown in Fig.
\ref{IIIstageP}, while the energy loss and energy spread are
plotted in Fig. \ref{IIIstageen}.

\begin{figure}
\includegraphics[width=1.0\textwidth]{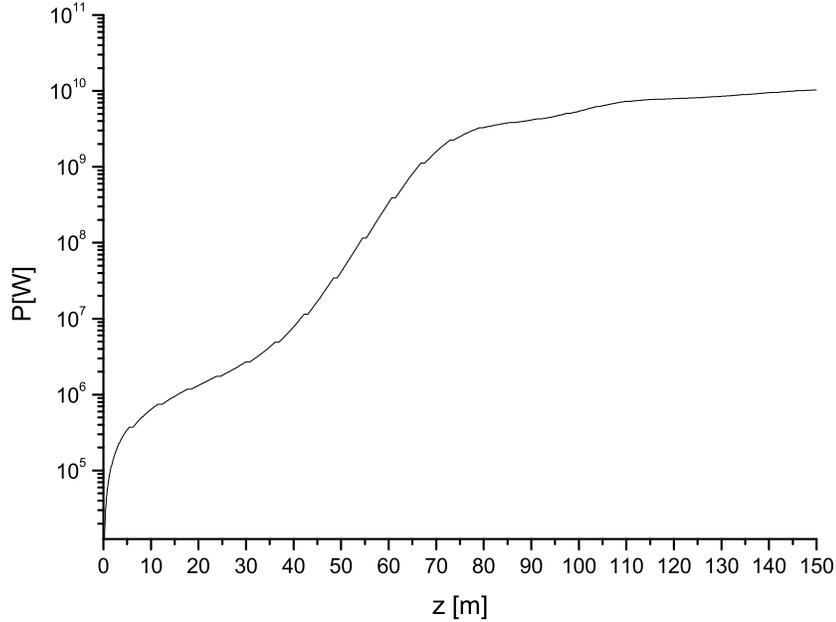}
\caption{Average output power from the third stage.\label{pave}}
\end{figure}

There are still about $100$ m of undulator available after the end
of the third stage, and in principle one may continue the SASE
process beyond saturation, as shown in Fig. \ref{pave}. However,
we choose to have the end of the third part of the undulator
coinciding with the output level of the second stage.

Inspection of Fig. \ref{IIIstageP} and Fig. \ref{IIIstageen} show
that the electron beam is now spoiled by the SASE process, i.e.
the fresh part of the bunch has lased.

\subsection{Results \label{result}}

\begin{figure}[tb]
\includegraphics[width=1.0\textwidth]{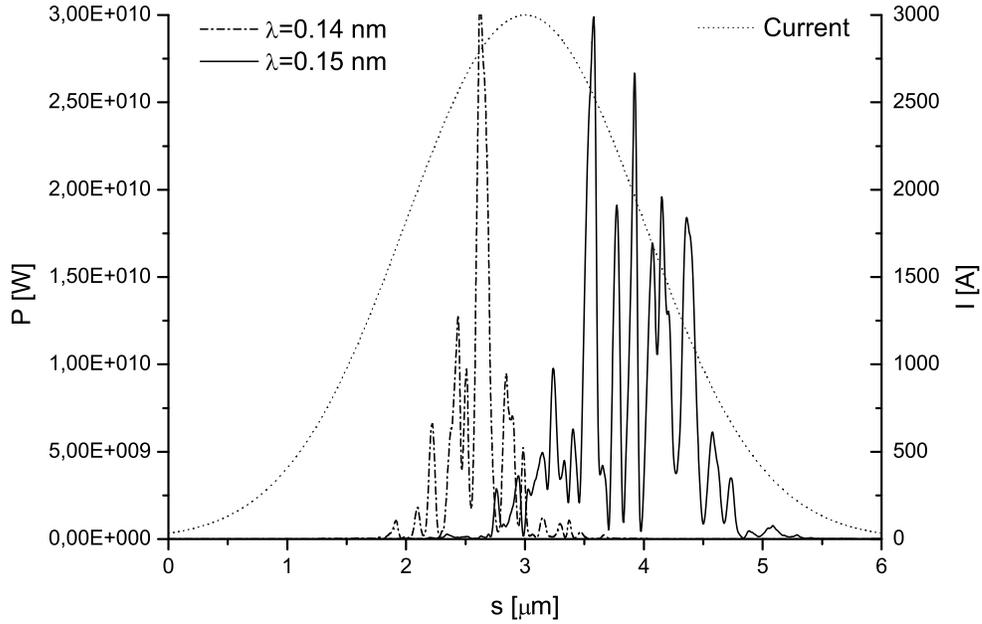}
\caption{Superimposed radiation pulses at $1.4{\AA}$ and $1.5
\AA$. The bunch current is also shown.} \label{pprobe}
\end{figure}

The overall result of our simulations is shown in Fig.
\ref{pprobe}. The output pulses from the second and the third part
of the undulator are shown, together with the current profile of
the electron bunch. Two femtosecond, ten GW level pulses of
coherent x-rays with two different colors can be easily generated
with this method.

\section{Photon beam manipulation \label{reldelay}}

\subsection{Tunable relative delay }

Once the result shown in Fig. \ref{pprobe} is established, one
faces the task of transport and utilization of the two radiation
pulses to the experimental station. While transport can be
performed with the same optics without problems, utilization for
pump-probe experiments implies the capability of separating and
delaying the two pulses of a given temporal amount at the
experimental station. Investigating this capability goes beyond
the scope of this paper.

\begin{figure}[tb]
\includegraphics[width=1.0\textwidth]{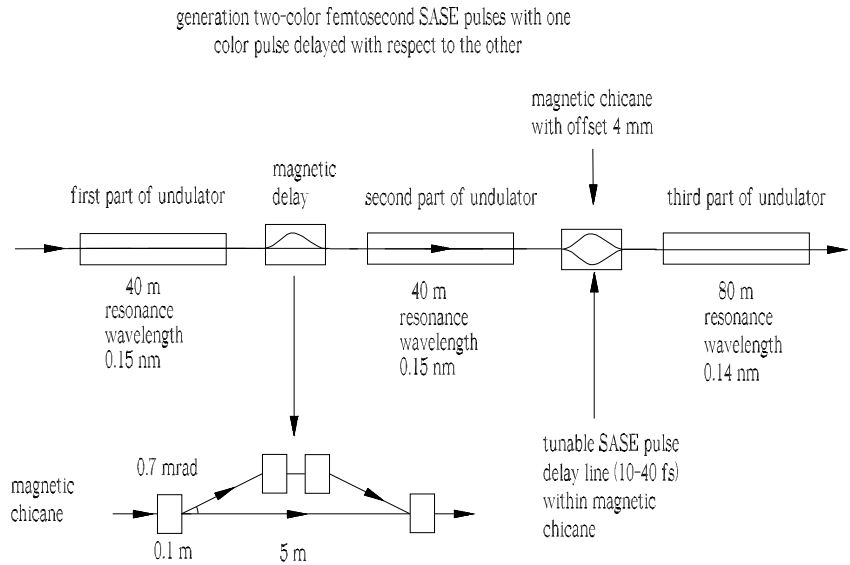}
\caption{Design of undulator system for short pulse (6 fs) mode
operation employing two color femtosecond pulses that are delayed
one with respect  to the other} \label{delay}
\end{figure}

\begin{figure}[tb]
\includegraphics[width=1.0\textwidth]{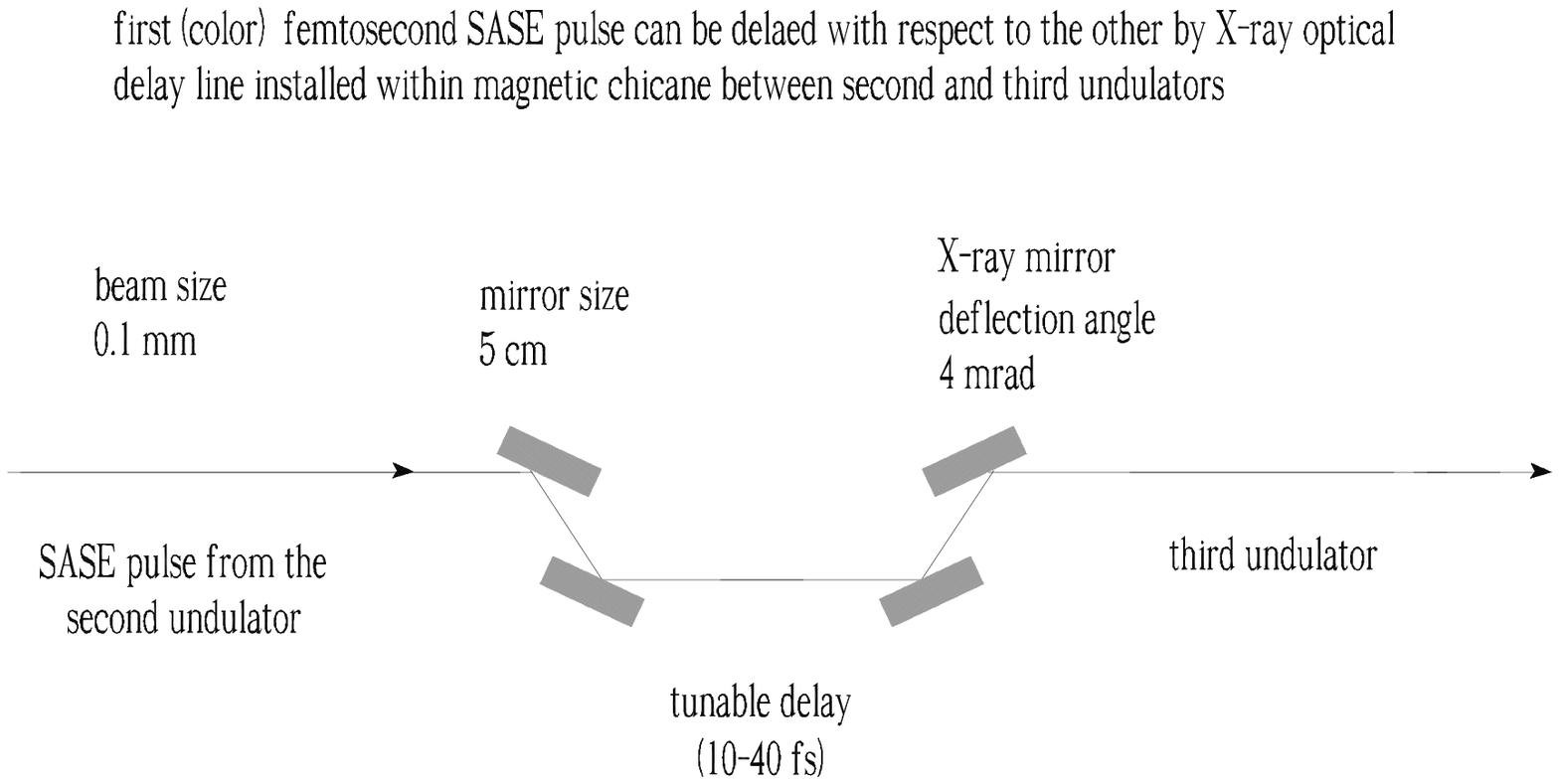}
\caption{Scheme for delaying the first (color) femtosecond SASE
pulse with respect to the other within undulator system. X-ray
optical system can be installed within magnetic chicane between
second and third undulator}\label{undudelay}
\end{figure}

A possible alternative is to include a tunable delay already at
the level of the SASE undulator setup. This can be done with usual
grazing-incidence optics. The idea is to install a mirror chicane
between the second and the third part of the undulator, as shown
in Fig. \ref{delay} and Fig. \ref{undudelay}. The function of the
mirror chicane is to delay the $0.15$ nm-radiation relatively to
the bunch and, therefore, also relatively to the $0.14$ nm-pulse.
The glancing angle of x-ray mirrors is as small as $2$ mrad.
Inside the photon-beam transport tunnel, the transverse size of
the radiation requires long mirrors, in the meter size. In
contrast to this, at the undulator location, the transverse size
of the photon beam is smaller than $100 ~\mu$m, meaning that the
mirror length would be just about $5$ cm. Moreover, the
short-pulse mode will relax the heat-loading issues.

Of course, in order to install the mirror chicane one needs to
first create an offset for the electron trajectory, meaning that a
magnetic chicane should be inserted at the position of the
mirror-chicane. The mirror chicane can be built in such a way to
obtain a delay of the SASE pulse of about $40$ fs. This is enough
to compensate a bunch delay of about $10$ fs from the magnetic
chicane, and to provide any desired temporal shift in the range
$0-30$ fs, as shown in Fig. \ref{delay}.

\subsection{Separating the two-color pulses}

\begin{figure}[tb]
\begin{center}
\includegraphics[width=100mm]{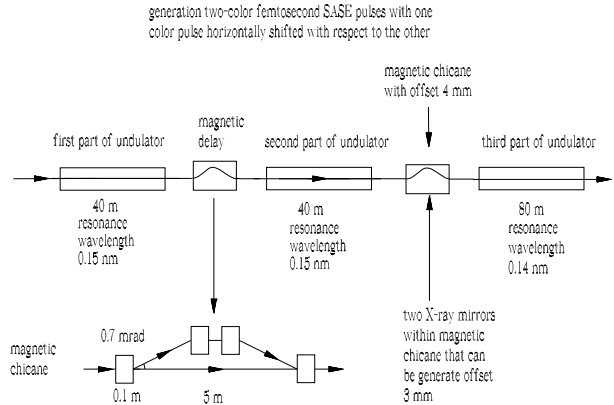}
\caption{\label{horsh1} Sketch of an undulator system for
short-pulse mode of operation employing two color femtosecond
pulses that are horizontally shifted one with respect to the
other. }
\end{center}
\end{figure}
\begin{figure}[tb]
\begin{center}
\includegraphics[width=100mm]{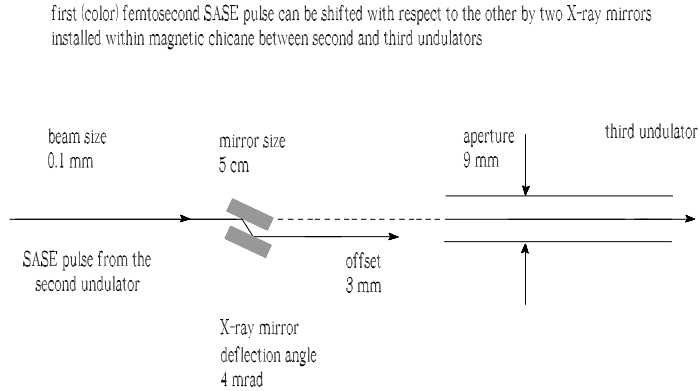}
\caption{\label{horsh2} Scheme for horizontally shifting the first
femtosecond SASE pulse with respect to the other within the
undulator system. Two X-ray mirrors can be installed within the
magnetic chicane between the second and the third undulator. }
\end{center}
\end{figure}
It should be noted that the main difficulty concerning the
manipulation of the two-color photon beam consists in the
separation of the two colors. Once this task is performed, the
delay problem can be easily solved in the experimental hall with
the help of mirrors. Of course, the two colors can be separated in
the experimental hall as well with the help of crystals. However,
this would lead to a loss of photons due to narrow bandwidth of
the crystals.

As an alternative to the tunable relative delay considered above,
here we propose to separate the two colors already in the
undulator with the help of x-ray mirrors. The idea is sketched in
Fig. \ref{horsh1} and Fig. \ref{horsh2}. The two colors can be
separated horizontally by two mirrors installed within the chicane
after the second undulator. The horizontal offset can be about $3$
mm, which is enough for separating the two-color pulses, because
at the position of the optical station the FWHM beam size is less
than a millimiter. Additionally, mirrors can also be used to
generate a few $\mu$rad deflection-angle, which is not important
within the undulator but will create further separation of a few
millimeters at the position of the experimental station.

\section{Long-pulse operation mode \label{long}}

\begin{figure}[tb]
\begin{center}
\includegraphics[width=100mm]{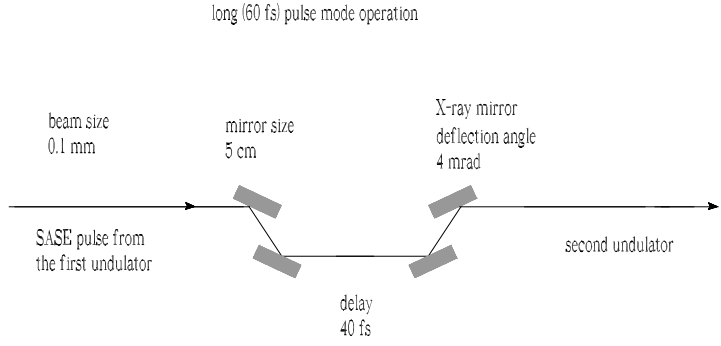}
\caption{\label{delayopt} X-ray optical system for delaying the
SASE pulse in the case of long (60 fs) pulse mode operation }
\end{center}
\end{figure}

\begin{figure}
\begin{center}
\includegraphics*[width=100mm]{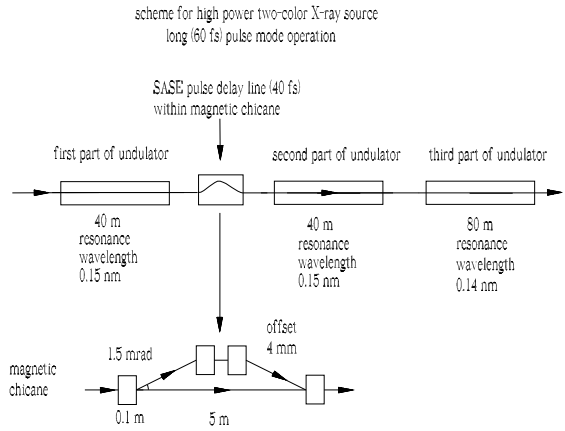}
\caption{\label{undusyslong} Design of undulator system for long
(60 fs) pulse mode operation.  }
\end{center}
\end{figure}

\begin{figure}
\begin{center}
\includegraphics*[width=100mm]{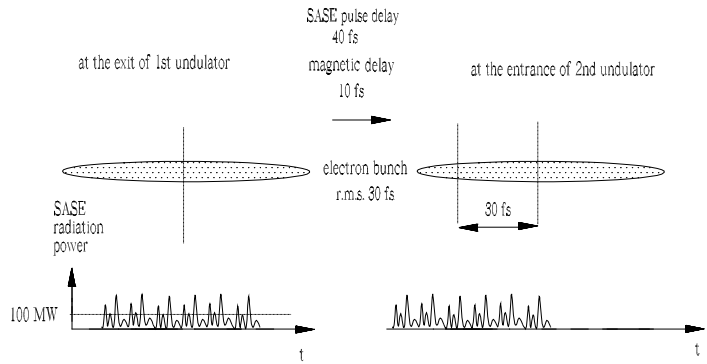}
\caption{\label{frelong} Sketch of principle of "fresh bunch"
technique for long (60 fs) pulse mode operation.  }
\end{center}
\end{figure}
As mentioned in the introduction, our method is also suitable for
the long-pulse operation mode. A straightforward way of
application consists in increasing the magnetic chicane angle up
to about $2.5$ mrad, thus providing a relative delay of the
electron beam with respect to the electron pulse of $30$ fs.

Alternatively, one may use an optical delay line which would,
however, require extra-hardware. A mirror chicane like that
considered in section \ref{reldelay} would delay the radiation
relatively to the bunch (see Fig. \ref{delayopt}). The difference
with respect to the optical chicane for obtaining a tunable delay,
considered above, is that now the optical delay line should be
installed between the first and the second undulator. As before,
the combination optical delay-magnetic chicane can be built in
such a way to obtain a delay of the SASE pulse relative to the
electron beam of about $30$ fs, as shown in Fig. \ref{frelong}. In
this way, the strength of the magnetic chicane can be reduced, and
the deflection angle of the electron beam can be set to $1.5$
mrad. Also note that heat-loading problems would be strongly
mitigated by the fact that the photon pulse is still in the linear
regime, at low power ($100$ MW).

\section{Conclusions}

We presented a method to obtain two short (sub-ten fs), powerful
(ten GW-level) pulses of coherent x-ray radiation at different
wavelengths for pump-probe experiments at XFELs. The idea is based
on the fresh-bunch technique \cite{HUAYU,SAL1,SAL2}, and can be
realized based on the ideal performance of the beam formation
system, which has been experimentally demonstrated at LCLS
\cite{LCLS1,LCLS2}.

The method relies on the substitution of a single undulator
segment with a delay stage. In its simplest form, that can be
implemented in the short-pulse operation mode demonstrated at LCLS
\cite{DING}, the delay stage is just constituted by a magnetic
chicane. The method can also be implemented in the long-pulse
operation mode, in which case an extra optical delay stage must be
provided in the same undulator segment. Manipulation of a single
segment returns great advantages, but it should be clear that our
technique is actually based on the experimentally demonstrated
availability of optimal full-scale facility parameters.

Since it is based on manipulation of a single segment, the method
is at the same time low cost, and very robust. Also, it does not
perturb the baseline long-pulse operation mode, and carries no
risks for the operation of the machine. Finally, and most
importantly, the two-color X-ray  pulses are precisely
synchronized at the femtosecond level, since they both are
produced by the same electron bunch, and there is no time jitter
between the pulses.

Our study is exemplified with parameters typical of the SASE2 line
of the European XFEL. Nevertheless it can be applied at any XFEL
facility.

\section{Acknowledgements}

We are grateful to Massimo Altarelli, Reinhard Brinkmann, Serguei
Molodtsov and Edgar Weckert for their support and their interest
during the compilation of this work.

\end{document}